\documentclass[a4paper]{article}

\usepackage{amssymb}
\usepackage{amsmath}
\usepackage{latexsym}

\def\span {\mbox{span}}
\def\setC{\mathbb{C}}
\def\R{{\rm I\!R}}

\def\bbN{\mathbb{N}}
\def\lr{Lachi\`eze-Rey}

\def\uc  {universal covering}
\newcommand{\ket}[1]{\mid #1 >}
\newcommand{\bra}[1]{<#1 \mid}

\def\calV {{\cal V}}
\def\calH {{\cal H}}
\def\calY {{\cal Y}}
\def\calF {{\cal F}}
\def\I{{\rm I\!I}}
\def\nn {{\bf n}}
\def\th {$^{th}$}

\def\circle {\circ}

\def\ON {orthonormal}
\def\CG {Clebsch-Gordan}

\def\ul{u_l}
\def\ur{u_r}
\newcommand{\bun}{(\mathcal{B}_1)}
\newcommand{\bdeux}{(\mathcal{B}_2)}

\def\llc{\mbox{[\hspace{-.13em}[}}
\def\lrc{\mbox{]\hspace{-.13em}]}}
\def\st{\mathcal{S}^3}

\newcommand{\mat}[4]{
\left(
\begin{array}{c c}
#1 & #2 \\
#3 & #4 \\
\end{array}
\right)
}

\title{Laplacian eigenmodes for  spherical spaces }
\author{M. Lachi{\`e}ze-Rey $^{1,2}$ and S. Caillerie $^{1}$\\
1. Service d'Astrophysique, C. E. Saclay\\
91191 Gif sur Yvette cedex, France\\
2. CNRS : UMR 7164}
\begin{document}
%      --------
%%%%%%%%%%%%%%%%%%%%%%%%%%%%%%%%%%%

\maketitle

\abstract{
The possibility that our space is multi - rather than singly - 
connected has gained a renewed interest after the discovery of the low 
power for the first multipoles of the CMB by WMAP.  To test  the 
possibility that our space is a multi-connected spherical space, it is 
necessary to know the eigenmodes of such spaces. Excepted for lens and 
prism space, and in some extent for  dodecahedral space, this remains 
an open problem. Here we derive  the eigenmodes of all spherical 
spaces. For dodecahedral space, the
demonstration is much shorter, and  the calculation   method  much 
simpler  than  before. We also apply to tetrahedric, octahedric and 
icosahedric spaces. This completes the knowledge of eigenmodes for 
spherical spaces, and opens the door to new observational tests of 
cosmic topology.

The vector space ${\cal V} ^{k }$ of the eigenfunctions of
the Laplacian on  the three-sphere   $\st$, corresponding to the same
eigenvalue  \mbox{$\lambda _{k} = -k~(k+2)$}, has dimension
$(k+1)^2$. We show  that the Wigner functions provide a basis for such 
space.
Using the properties of the latter, we express the behavior of a 
general function of ${\cal V} ^{k }$ under    an arbitrary  rotation 
$G$  of SO(4).
   This offers the possibility to select those functions of ${\cal V} ^{k
}$  which remain invariant   under $G$.

Specifying $G$ to be a  generator  of the holonomy group of a spherical 
space $X$, we   give   the expression of the
  vector space ${\cal V}_X ^{k }$ of the eigenfunctions of $X$. We 
provide a method  to calculate the eigenmodes up to arbitrary order. As 
an illustration, we give the first  modes for the spherical spaces 
mentioned.
}

%%%%%%%%%%%%%%%%%%%%%%%%%%%%%%%%%%%
%
\section{Introduction}
%        ------------
%%%%%%%%%%%%%%%%%%%%%%%%%%%%%%%%%%%

\subsection{Cosmic Topology}

The possibility that  spatial sections of our Universe have a
multi-connected topology offers   interesting alternatives to the
standard big bang models \cite{lalu}. The  recent   suggestion
\cite{Luminet}  that a
peculiar variant, namely the Poincar\'e dodecahedral space, may
explain the deficiency in the first modes of the angular  power
spectrum of the CMB, as observed by WMAP has renewed the interest to
this question. It is presently an important task of cosmology to check
this possibility.

The best hope to test the possible   multi-connectedness of space
comes from the CMB fluctuations. For this task, the most important
characteristic of a multi-connected space $X$ is
the set of its   eigenmodes, a subspace of the  modes of its  \uc 
~\cite{Riazuelo}.
The main consequences on the CMB are (i) a suppression of power at the
scales comparable to, or larger than,  the circumference of $X$, and
(ii) a violation of
isotropy at these scales. Thus, observational signatures may occur  as
a reduction of power in the angular CMB spectrum (as reported by
\cite{Luminet}), and as the imprint of correlations between different
modes.
The predictions of these  observable effects  require  the knowledge
of  the eigenmodes [of the Laplacian]  of $X$. For instance,  \cite{Luminet} use a limited
number of eigenmodes of the Poincar\'e space, calculated
numerically. More precise checks  require the knowledge of a larger
number of modes of this space.  The examination of other models require
the
knowledge of the eigenmodes of other spherical spaces.

This paper is devoted to the calculation of the eigenmodes of   all
variants of the
multi-connected spherical spaces.  They are  defined as
quotients  $X\equiv \st/\Gamma$, where the three-sphere $\st$ is
the \uc ~space of $X$, and  $\Gamma
     \subset  $SO(4)  is the holonomy
group of $X$.  The  transformations of $\Gamma$  leave $X$   invariant.
The   compactness of $\st$ implies the   finiteness of  $\Gamma$. The
latter is   finitely generated, {\it i.e.} by a finite
number of transformations.

\subsection{Spherical spaces}

In the  family of spherical spaces, the eigenmodes of   lens and  prism
spaces have been calculated by
\cite{Leh}, \cite{Leh2} (see also \cite{mlrS3}) in analytic form.  For
the  dodecahedral space, only the very first modes  had been
calculated numerically, by  \cite{Luminet}. Later,
\cite{mlrDodec}~reduced   the  calculation to an eigenvalue problem
of reduced dimensionality,  allowing their estimations   up to
arbitrary order.
       For the other  spherical spaces, the knowledge  of the eigenmodes
remains an open problem. Here, we present a general method to
calculate the eigenmodes of any of the spherical spaces $X=\st
/\Gamma$.

       For a given eigenvalue $\lambda _k$, the eigenmodes of $X$
describe an eigenspace~${\cal V}_X^{k}$. The  dimension of this
vector space has  been calculated, very generally,  by \cite{Ikeda}.
We use his results as a cross-check of the calculations below.

    Our  calculation is based on the fact that each $\Gamma$ has two
generators $G_\pm$, which are single left action rotations of SO(4).
We obtain the  straightforward decomposition
$${\cal V}_X^{k} = {\cal V}_{G_+}^{k} \cap {\cal V}_{G_-}^{k},$$
and we give  explicit analytic expressions of the $ {\cal
V}_{G_\pm}^{k}$, which are the vector spaces of the eigenfunctions of
$\st$ invariant under $G_\pm$.

Although the expressions of the two $ {\cal V}_{G_\pm}^{k}$'s are very
simple,  there is no general  analytic expression of  their
intersection.
Thus, to obtain an explicit expression of the modes, we propose a
general method, which applies to any  spherical space. It is based on
the fact that a basis of  ${\cal V} ^{k}$ is provided by the Wigner
functions. From  the rotation properties of the latter, we derive a
simple invariance condition of the modes of $\st$ under $G_\pm$, and
thus under $\Gamma$.
This provides  the eigenmodes of $X$, which are the   eigenmodes of
$\st$ which remain invariant under  $\Gamma $.

In (\ref{S3}), we analyze the eigenfunctions of $\st$. We prove that
the Wigner functions provide a convenient basis $\bdeux$, different
from the widely used basis $\bun$ provided by the usual hyperspherical
harmonics.
In (\ref{rotation}), we derive the rotation properties of this basis
under SO(4), from which we deduce
those of all  eigenfunctions of $\st$. This allows to give a criterion
for the invariance of such a function under an arbitrary rotation of
SO(4).
In (\ref{S3Gamma}), we give  an explicit analytic expression for the
invariant subspaces $ {\cal V}_{G_\pm}^{k}$, and we provide a general
procedure to calculate the eigenmodes of any spherical space. The first
modes are explicitly given in Appendix.

%%%%%%%%%%%%%%%%%%%%%%%%%%%%%%%%%%%
%
\section{Eigenmodes of $\st$}\label{S3}
%        ------------------
%%%%%%%%%%%%%%%%%%%%%%%%%%%%%%%%%%%

      The eigenmodes of $\st$ are the eigenfunctions of the Laplacian
$\Delta_{\st}$ of $\st$.
The eigenvalues   are
known to be of the form $\lambda_{k} = -k(k+2)$, where $k \in \bbN$.
The eigenmodes corresponding to  a given value of $k$   span the
eigenspace~${\cal V}^{k}$, a vector space of dimension $(k+1)^2$.
They are the   solutions of the Helmoltz equation
\begin{equation} \label{Helmoltz}
\Delta_{\st} f =\lambda _k f= -k(k +2)\ f, \quad k \in
\bbN.\end{equation}
Each  eigenspace ${\cal V}^{k}$ realizes   the
$(k+1)^2$ dimensional irreducible representation of SO(4), the
isometry group of $\st$.

     The corresponding eigenspace ${\cal V}_X^{k}$, of $X = \st/\Gamma$,
the goal of our search, is the vector space of functions of  ${\cal
V}^{k}$ which are $\Gamma$ - invariant.
Each ${\cal V}_X^{k}$ is a subvector space of ${\cal V} ^{k}$, whose
dimension (possibly zero) is the degeneracy of $\lambda_{k}$ on $X$.

\vspace*{0.3cm}
\subsection{Basis $\bun$: hyperspherical harmonics}

Let us call $\bun$ the most  widely used (\ON) basis for ${\cal V}^{k}$
provided by  the hyperspherical  harmonics
\begin{equation}
\label{ }
\bun \equiv \{\mathcal{Y}_{k \ell m},\ 0 \leq \ell\leq k , -\ell \leq
m\leq \ell \}
\end{equation}
It generalizes the usual spherical  harmonics $Y_{\ell  m}$ on the
sphere (note that $ {\cal Y}_{k \ell  m}  \propto Y_{\ell  m}$). In
fact, it can be shown (\cite{Bander}, \cite{erd} p.240,\cite{fry})
that a
basis of this type exists on any sphere $S^{n}$. Moreover,
\cite{erd} \cite{fry}  show  that the $\bun$  basis  for $S^n$  is
``naturally generated'' by the  $\bun$ basis  for $S^{n-1}$. In this
sense,
the $\bun$  basis  for $\st$ is  generated by the  usual spherical
harmonics $Y_{\ell m }$ on the 2-sphere $S^2$.

The generation process involves harmonic polynomials constructed
from  null complex vectors. The basis $\bun$ is in fact based on the
reduction of the representation of SO(4) to representations of
SO(3): each ${\cal Y}_{k \ell m}$ is an eigenfunction of an SO(3)
subgroup of SO(4) which leaves a selected point of $\st$
invariant. This makes these functions useful when one considers the
action of that particular SO(3) subgroup. But they show no simple
behaviour under a general rotation.

This basis is adaptated to the usual polar spherical coordinates
$(\psi,\theta,\phi)$.

%%%%%%%
\vspace*{0.3cm}
\subsection{The parabolic basis $\bdeux$}

The second (\ON) basis of ${\cal V}^k$   is specific to $\st$:
\begin{equation}
\label{B2}
\bdeux \equiv \{T_{k;m_1,m_2},\ -k/2\leq m_1, m_2\leq k/2\}
\end{equation}
where $m_1$ and $m_2$ vary  independently by integers  (and,
thus, take integral  or semi-integral  values according to the parity
of $k$). It is for instance introduced in \cite{Bander} by group
theoretical arguments, it appears naturally  adapted to the systems of
toroidal coordinates to describe $\st$:
\begin{displaymath}
\left\{
\begin{array}{ccc}
x^0 & = & r~\cos \chi ~\cos \theta \\
x^1 & = & r~\sin \chi ~\cos \phi   \\
x^2 & = & r~\sin \chi ~ \sin \phi  \\
x^3 & = & r~\cos \chi ~\sin \theta \\
\end{array}
\right.
\end{displaymath}
spanning  the  range $0 \le \chi \le \pi/2$, $0 \le \theta \le 2\pi$
and $0 \le \phi \le 2\pi$. Here, the $(x^{\mu})$ represent the
cartesian coordinates of the embedding space $\R ^4$ (see \cite{mlrS3},
\cite{Leh}), and as we consider only spherical spaces, $r$ is thus the
radius of $\st$, which is a constant assumed hereafter to be 1.

The explicit expression of the $T_{k;m_1,m_2}$ can be found in
(\cite{Bander}, \cite{Leh}, see \cite{mlrS3}). It is proportional to a
Jacobi
polynomial:
\begin{equation}
\label{B2modes}
T_{k;m_1,m_2} (X) = C_{k;m_1,m_2}~ (\cos \chi ~ e^{i \theta})^{\ell
} ~(\sin \chi ~ e^{i \phi}) ^{m} ~P^{( m,\ell)}_d(\cos (2 \chi)),
\end{equation}
where we have written
$$\ell = m_1+m_2 , ~m = m_2-m_1,~d = k/2-m_2.$$  The factor
$C_{k;m_1,m_2}$ can be computed from
normalization requirements (there is a factor $\sqrt{2}$ lacking in
\cite{mlrS3} formula) as
\begin{equation}
C_{k;m_1,m_2} \equiv \sqrt{\frac{k+1}{2\pi^2}}
\sqrt{\frac{(k/2+m_2)!(k/2-m_2)!}{(k/2+m_1)!(k/2-m_1)!}}.\end{equation}

We  also note these useful proportionality relations:
\begin{equation}
\hspace*{-0.2cm}
\begin{array}{l c l}
\cos^{\ell }\chi ~ \sin^{m}\chi  ~P^{( m,
\ell)}_{\frac{k-\ell-m}{2}}(\cos 2 \chi)
     & \hspace*{-0.25cm}\propto\hspace*{-0.25cm} & \cos^{\ell} \chi ~ 
\sin
^{ -m} \chi  ~P^{( -m,
\ell)}_{\frac{k-\ell+m}{2}}(\cos 2 \chi) \\
\rule{0pt}{0.6cm} &\hspace*{-0.25cm} \propto\hspace*{-0.25cm} &
\cos^{-\ell} \chi~ \sin^{m}\chi ~P^{(m,
-\ell)}_{\frac{k+\ell-m}{2}}(\cos 2 \chi).
\end{array}
\end{equation}

Since $ \bdeux$ is a basis, we have $\calV ^k=\span\left(
T_{k;m_1,m_2}    \right)_{-k/2\leq m_1, m_2\leq k/2 }$.
It appears convenient to  define the subvector spaces
\begin{equation} \label{Vkm}
    \calV ^{k,m_1} \equiv \span\left(  T_{k;m_1,m_2}  \right)_{-k/2\leq
m_2\leq k/2 }.
\end{equation} We have
$$\calV ^{k}= \bigoplus  _{m_1 = -k/2}^{k/2}\calV ^{k,m_1}.$$

We show now  that  the $T_{k;m_1,m_2}$  identify, up to a constant, to
the Wigner D-functions.

\vspace*{0.3cm}
\subsection{The Wigner D-functions}

      The Wigner D-functions are defined as functions on the Lie group
SO(3):
\begin{equation}
\label{Wignerdef}
\begin{array}{c c l}
\textrm{SO(3)} & \longrightarrow & \mathbb{C} \\
g & \longmapsto & {\cal D}_{m' m}^j (g) \equiv \bra{jm'} R_g  \ket{jm}.
\end{array}
\end{equation}
Here, $g$ is rotation of SO(3), and  $R_g$ expresses its  natural left
action $f \mapsto R_g f$  on functions:
$R_g f(x) \equiv f(g^{-1}x)$.
The vectors $\ket{jm}$,  $m\in \llc -j,j\lrc$ form an orthonormal
basis for $\calH ^{j}$, the $2j+1$ dimensional irreducible
representation of SU(2). When $j$ is integer, this is an IUR
of SO(3), and the  $\ket{jm}$ can  be taken as the usual spherical
harmonics $Y_{\ell m}$.  This implies
\begin{equation}
\label{ }
R_g Y_{j m} = \sum_{m'}  {\cal D}_{m' m}^{j} (g^{-1})~ Y_{jm'}
\end{equation}
(Care must be taken that one finds in the literature other definitions
with indices exchanged; we adopt here the notations of \cite{Edmonds}).

As manifolds, the three-sphere   $\st$ and SU(2) are
identical. This  allows to identify any point of $\st$ to an element
of SU(2) by
the following relation:
\begin{equation}
\label{ }
\begin{array}{c c c}
\st & \longrightarrow & \textrm{SU(2)} \\
x \equiv (\chi,\theta,\phi) & \longmapsto & u_x\equiv
\mat{\cos\chi~e^{i\theta}}{i~\sin\chi~e^{i\phi}}{i~
\sin\chi~e^{-i\phi}}{\cos\chi~e^{-i\theta}} \\
\end{array}.
\end{equation}
(Note that one finds other phase choices for this identification in
the literature).

On the other hand,
there is a group  isomorphism between SU(2) and SO(3)/$\mathbb{Z}_2$,
where
$\mathbb{Z}_2$ refers to the multiplicative group $\{-1,1\}$.  Thus,
each element $u$ of SU(2) defines an SO(3) rotation $g_u$. In
practice, a   rotation of SO(3) is parametrized by its Euler angles
$\alpha$, $\beta$, $\gamma$. Taking into account the identification
above, the correspondence takes the form:

      \begin{eqnarray}
     SU(2) & \mapsto  & SO(3)\\
u = (\chi,\theta,\phi)  & \mapsto  & g_u=(\alpha ,  \beta ,  \gamma),
\end{eqnarray}
where \begin{equation}
\chi = \frac{\beta}{2} \qquad \theta = \frac{\alpha+\gamma}{2} \qquad
\phi = \frac{\alpha-\gamma}{2}.\end{equation}

This allows to consider the Wigner D-functions as functions on SU(2)
and, thus, on $\st$. Their explicit expression (see for instance
\cite{Edmonds}) shows their identification to the previous functions
$T_{k;m_1,m_2}$:
\begin{equation}
\label{TD}
D _{m_2,m_1}^{k/2}(u_x) \equiv
\sqrt{\frac{2\pi^2}{k+1}}~T_{k;m_1,m_2}(u),
\end{equation}
with $-k/2\le m_1,m_2 \le k/2$.
This will allow to use their very convenient properties with respect
to rotations.

\subsection{Change of basis}

We have two bases for $\mathcal{V}^k$. Calculations may be more
convenient in one or the other. Most numerical codes use a development
in the basis $\bun$. Here we will calculate the eigenmodes of $X$ in
the basis  $\bdeux$. A conversion requires the formulae for the change
of basis, which result from the properties of the Wigner functions.

The expansion may be found in  \cite{Gazeau} (see also\cite{Bander},
which uses however different conventions):
\begin{equation}
\label{ }
\mathcal{Y}_{klm} (u)= i^{k+l-m} \sum_{m_1,m_2}
(-1)^{m_2}(\frac{k}{2},-m_1;\frac{k}{2},m_2 \mid l,m)~
T _{k;m_1,m_2}(u),
\end{equation}
where the $(j,m_1;j,m_2 \mid l,m)$ are the \CG ~coefficients.
We can invert this formula by multiplying it by another Clebsch-Gordan
coefficient $(j,m_3;j,m_4 \mid l,m)$ and by summing on the 2 indices
$l$ and $m$. The orthonormality of the Clebsch-Gordan coefficients
then gives the relation (after renaming indices):
\begin{equation}
\label{ }
T _{k;m_1,m_2}(u) = (-1)^{m_2} \sum_{l,m}
i^{-k-l+m} (\frac{k}{2},-m_1;\frac{k}{2},m_2 \mid
l,m)~\mathcal{Y}_{klm}(u)
\end{equation}

The properties of the \CG ~coefficients (non zero only when
$-m_1+m_2=m$ with the previous notation)  reduce these formulae,
respectively, to
\begin{equation}
\label{ }
\mathcal{Y}_{klm} (u) = \sum_{m_1} i^{k+l+m_1+m_2}
(\frac{k}{2},-m_1;\frac{k}{2},m_1+m \mid l,m)~
T _{k;m_1,m_1+m}(u).
\end{equation}
and
\begin{equation}
\label{ }
T _{k;m_1,m_2}(u) = \sum_{l} (\frac{k}{2},-m_1;\frac{k}{2},m_2 \mid l,
m_2-m_1) ~\frac{\mathcal{Y}_{kl (m_2-m_1)} (u)}{i^{k+l+m_1+m_2}}.
\end{equation}

\section{SO(4) rotations}\label{rotation}

\subsection{General case}

The isometries of $\st$ form the group SO(4), the rotation group of
the embedding space $\R ^4$.
We note $G:x \mapsto Gx$ such an isometry, and define its action on a
function $f$ of $S^3$  as
$$R_G:~f \mapsto R_G f; ~R_G f(x) \equiv f(G^{-1} x).$$

Using the previous identification between $\st$ and SU(2), it is very
convenient to use the   SU(2)$\times$SU(2) representation of SO(4).
Any rotation $G$ of SO(4) is represented  (modulo a sign) as a    pair
$ (\ul,\ur)$ of two SU(2) matrices, respectively acting on the left
and on the right on  $u_x$,   an element of SU(2):  
\begin{equation}
\label{rSO4}
\begin{array}{c c c l}
(\ul,\ur): &   SU(2) & \longrightarrow &  SU(2) \\
&  u_x & \longmapsto & \ul\cdot u_x \cdot\ur =u_{Gx},
\end{array} \end{equation}
where all matrices belong to SU(2). (The SO(4) matrices of
the form  $(\ul,\ur=\ul^{-1})$ form a specific SO(3) subgroup).

The resolution of identity in $\calH ^{j}$ (or equivalently, the
composition of two successive SO(3)  rotations, with $R_{g_1 \circle
g_2}= R_{g_2} \circle R_{g_1})$, implies the well known  addition
formula
\begin{equation}
\label{additionD}
D_{m m'}^j (u  v)=\sum _{m''}~D_{m m''}^j (u)~
D_{m'' m'}^j (v).
\end{equation}
Using (\ref{TD}), this   implies
\begin{equation}
\label{additionT}
T_{k;m 'm}  (x_{u v})=\sum _{m''}
D_{m m''}^{k/2} (u  ) ~T_{k;m' m''  }  (x_v  ).
\end{equation} Here, $x_u$ is the point of $\st$ corresponding to
$u\in $ SU(2).
Now we will    use the same  notation  for a function on $S^3$ and on
SU(2): $f(x_u) = f(u)$.

In these formulae, $u v$ refers to the  product of two SU(2)
matrices. We are now  able to compute the action of a rotation $G$ of
SO(4) on the basis  functions $T_{k;m_1,m_2}$ by using the
decomposition
(\ref{rSO4}) of $G$ and by using twice the addition formula
(\ref{additionT}) for $T_{k;m_1,m_2}$.

Thus, we obtain the rotation properties of the basis $\bdeux$ of
eigenfunctions under the  rotation $G \approx (u_l,u_r)$:
\begin{equation}
\label{ }
\begin{array}{c c l}
R_G T_{k;m_1,m_2} (v) \! &\! \equiv\! &\! T_{k;m_1,m_2} (\ul^{-1}
~v~\ur^{-1}) \\
\rule{0pt}{16pt} \!& \!=\! & \!\displaystyle{\sum_{m}\sum_{m'}\ D_{
m_2 m}^{k/2} (\ul^{-1})\ D_{m' m_1}^{k/2} (\ur^{-1})\ T_{k;m'
m}(v)}.
\end{array}
\end{equation}

This completely specifies the rotation properties of the basis $\bdeux$
under SO(4).

\subsection{Single action rotations}

A special case  is a single left action rotation  (hereafter SLAR), $G
\equiv   (u_l, u_r\equiv \I _{SU(2)})$, where  $u_l \in$SU(2) defines
an SO(3)
rotation $g$. The previous    formula reduces to:
\begin{equation} \label{rotT}
\begin{array}{c c l}
R_G T_{k;M,m_2} (v)\! &\! \equiv\! & \! T_{k;m_1,m_2} (\ul^{-1}~v) \\
\rule{0pt}{16pt}\! & \!=\! &\! \displaystyle{\sum _{m} ~ D_{m_2 m}^{k/2}
(\ul^{-1}) ~T_{k;M m}(v)} .\\
\end{array} \end{equation}

We  note  that $G$ preserves  the first  index $M=m_1$. This means
that each ${\cal V}^{k,M}$ (see \ref{Vkm}) is invariant under $G$.
This implies that the research of  eigenfunctions invariant under a SLAR
$G$ may be performed  in each ${\cal V}^{k,M}$ separately. Moreover,
since the  invariance equation   does not depend on $M=m_1$, the
solution   in one of them  will give the solution for all of them.

> From (\ref{rotT}), the invariance condition reads
\begin{equation}
\label{rotT2} T_{k;M,m_2} (v) =\sum _{m} ~ D_{m_2 m}^{k/2}
(\ul^{-1}) ~T_{k;M m}(v).
\end{equation}

%%%%%%%%%%%%%%
\subsection{Invariant functions}

      The eigenfunctions of $X \equiv \st/\Gamma$ are
the eigenfunctions of $\st$ which remain invariant under $\Gamma$. A
necessary and sufficient condition is that they remain invariant under
the generators of $\Gamma$, which  are SLARs.

Thus, we search the eigenfunctions invariant under an arbitrary   
SLAR~$G$. For a given value   $k$, let us call their space $\calV_G 
^{k}
\subset
\calV ^{k} $. Since  $G$ preserves the vector space $\calV  ^{k,M }$,
we have the decomposition
$$\calV_G ^{k}= \bigoplus  _{M  = -k/2}^{k/2}\calV _G^{k,M };
~\calV _G^{k,M } \equiv \calV  ^{k,M } \cap \calV _G^{k }.$$

The  search of   the $G$-invariant  eigenmodes is equivalent to the
search of the  vector spaces $\calV_G ^{k,M}$, for the different
values of $k,M$, in $\calV ^{k,M}$.

In $\calV ^{k,M}$, the expansion of a function $f$ on the   canonical
basis, $$f = \sum _{m=-k/2}^{k/2} f_{k;Mm} ~T_{k;M,m},$$  describes it
as the vector $F$  with the    $ k+1$    components  $F_m \equiv
f_{k;Mm}$. Similarly, we write  for simplicity   the matrix $M_{mn}
\equiv D_{nm}^{k/2}(u_\ell ^{-1})$. Then,  equ.(\ref{rotT2}) takes the very simple
form
    \begin{equation} \label{eigen}
    F_m = \sum _n M_{m n}~F_n.
\end{equation}  This is
    an eigenvalue equation
$M F=F.$ Its solutions are the eigenvectors of the matrix $M$, with
the eigenvalue 1.

We  recall  (see equ.\ref{Wignerdef})  that  the Wigner function  are
the rotation matrix elements in the (k+1) - dimensional representation
$k/2$. Assuming
    $k=2 \ell $ even,  the vector $F$ represents also an  harmonic
function on the 2-sphere
$\psi _V  \equiv \sum _m F_m~Y_{\ell m} \in \calV ^\ell $.
Thus, (\ref{eigen}) becomes
$$\psi_V = ( R_g)^t~\psi_V  $$ (the subscript $t$ holds for matrix
transposition).  This is the condition  for  the function $f_V$ to be
invariant under the   SO(3) rotation $g^{-1}$. The solution of this
problem is well  known.
Let us call $\nn$ and $2\pi /N$ the oriented (unit) axis and angle
(assumed an integer divisor of $2 \pi$)  of  the rotation $g$.
\begin{itemize}
     \item {\bf Diagonal case:} When $\nn$ is along Oz, the rotation is
diagonalized : the SU(2) matrix,   $u=   \textrm{diag}(e^{i \pi /N
},e^{-i \pi /N})$,   $[R_g]_{mn} =\delta
_{mn}~e^{2i\pi ~m/N}$ and,  using   (\ref{B2modes}) and  (\ref{TD}),
\begin{equation}
\label{diagm}
\begin{array}{r c l}
D_{m_2 m}^{k/2} (u^{-1}) & = & \delta_{m
m_2}~(e^{-i\pi /N})^{m+m_2} ~P^{(m_2-m,m+m_2)}_{k/2-m}(1) \\
& = & e^{-2i\pi m/N } ~\delta_{m m_2}. \\
\end{array}
\end{equation}

The eigenspace corresponding to the eigenvalue 1 is thus  $\span (e_{N
\mu})_{-J_N\le \mu  \le  J_N}$,
     after defining $J_N \equiv  \lfloor k/2N\rfloor   $ ($\lfloor \cdot
\rfloor $ means integer part), and
    where $e_i$ is the i\th ~basis vector. In other  words, the      non
zero components of an  invariant vector have as index   an integer
multiple of $N$. The dimension of this space is $2 J_N+1$.

Coming back to SO(4), this results implies immediately that
$$\calV _G^{k,M}  = \span (T_{k;M,N \mu }),~-J_N\le \mu  \le  J_N .$$

     \item
When $\nn$ is not  along Oz,  we may diagonalize $R_g$ as $g =\rho
^{-1} ~D~ \rho$.
The eigenvalue equation becomes
$$f_V = \rho ^{-1} ~D~ \rho~f_V    .$$ This is the same eigenvalue
equation  as above,
    with the same solution, although not for $V$ but for $\rho V$.
The eigenspace corresponding to the eigenvalue 1 is thus  $\span (\rho
^{-1}~e_{N \mu})_{-J_N \le \mu \le J_N}$.

Coming back to SO(4), this results implies immediately that
$$\calV _G^{k,M}  = \span (\rho ^{-1}~T_{k;M,N \mu}), ~-J_N \le \mu
\le J_N .$$
\end{itemize}

\section{Eigenmodes of spherical spaces} \label{S3Gamma}

When we have two generators $G_\pm$, the eigenspace is simply
$$\calV _X^{k,M}= \calV _{G_+}^{k,M}  \cap \calV _{G_-}^{k,M} .$$

In  any spherical space,  we chose a frame   where one of the
generators of~$\Gamma$ is diagonalized, so that the invariance
condition relative to this generator takes the simple form
(\ref{mcdiag}). Thus, the problem is reduced to the search of the
invariance condition relative to the other generators.

\subsection{Diagonal matrices}

Applying  equation (\ref{rotT}) to a diagonal matrix as above, we obtain:
\begin{equation}
R_g T_{k;m_1,m_2} (v) = e^{-2i\pi ~m_2/N}~T_{k;m_1 m_2}(v)
\end{equation}

Let us assume that a    generator of the holonomy group of a spherical
space $X$ is represented by a diagonal left action SU(2) matrix $g
\approx (u_l,\I)$ with
$u_l$ of the diagonal  from above.
The invariance condition, under $g$, of the   basis functions for
this space takes the form \begin{equation}
T_{k;m_1,m_2} (u_l) = e^{-2i\pi m_2/N}T_{k;m_1 m_2}(u_l).
\end{equation}
Its solution is
\begin{equation}
\label{mcdiag}
m_2 = 0\pmod{N},\end{equation}
or $m_2 =N \mu$, where $\mu$ varies in the range $[-J_N..J_N]$.

    Hereafter, all sums involving $\mu,\nu$ are assumed to go in this
range ($2J_N +1$ values).
This is the selection rule for the eigenmodes invariant under $g$.

\subsection{A general procedure}

The spherical spaces considered below all have two generators which
are SLARs, $G_\pm \equiv (\gamma _\pm,\I)$.  Let us chose a frame
where $G_+$ is diagonal. Then,  form above, we can write
$$\calV _G^{k,M}  = \span (T_{k;M,N \mu}), ~-J_N \le \mu \le J_N.$$
In other words, all eigenfunctions $T_{k;M,m}$, with $m=N\mu$ are
$G_+$-invariant.

Since there is no simple way to calculate the intersection
    $$\calV _X^{k,M}= \calV _{G_+}^{k,M}  \cap \calV _{G_-}^{k,M} ,$$
we propose a practical procedure.

Let us expand a  $G _+$-invariant
function    in the basis $\bdeux$ as
$$f =\sum_{m}\sum_{\mu}f_{k;m,N\mu}~T_{k;m,N\mu} , ~~ \mu = -
J_N\cdot \cdot J_N  .$$

The invariance condition under the rotation $\gamma _{-}$, for such a
function, takes the form:
\begin{equation}
\begin{array}{c}
R_{G _{-}} f  = f : \\
\rule{0pt}{5ex}\displaystyle{
\sum_{m}\sum_{\mu}f_{k;m,N\mu}~
R_{G _{-}}T_{k;m,N\mu}(x) =
\sum_{m}\sum_{\mu}f_{k;m,N\mu}~T_{k;m,N\mu}(x)} .\\
\end{array}\end{equation}
With   the transformation law (\ref{rotT}), this becomes:
\begin{equation}
\label{condfk}
f_{k;m,N\mu} = \sum_{\nu} f_{k;m,N\nu}
~D^{k/2}_{N\nu,N\mu}\left( (\gamma_-) ^{-1}\right).\end{equation}
      This equation is independant of index $m$.

      This implies that    the eigenspace for $k$ splits  as the direct
sum
      $${\cal V}_X^k=\bigoplus _{m}~
    {\cal V}_X^{k,m}.$$
The degeneracy of the eigenvalue $\lambda _k$ is the dimension of
     ${\cal V}_X^k$. This implies that $$dim \left(  {\cal V}_X^k\right)
=(k+1)~
dim \left(  {\cal V}_X^{k,m}\right) .$$

       Thus, the solution for $m$ does not depend on $m$, and we can 
write
$f_{k;m,n} = f_{k;m_0,n}  $, where $m_0$ is some index in
$\llc -k/2;k/2\lrc$.
Equ.  (\ref{condfk})  is still an eigenvalue
equation, but in a vector space of restricted dimension $2 J_N+1$:
Let us define the matrix  ${\cal M}_k$, of order $2J_N+1$,  through
its   coefficients:
$$[{\cal M}_k]_{\mu,\nu} \equiv  D^{k/2}_{N\nu,N\mu}\left( (\gamma
_-)^{-1}\right), ~ \mu,\nu = -J_N  \cdot \cdot J_N  .$$ Note that
this matrix does not depend on the index   $M$. Let us also define the
$(2J_N+1)$- vector  ${\cal F}$ through its components :$$[{\cal
F}]_\mu \equiv f_{k;M,N\mu}.$$  Equation (\ref{condfk}) could thus be
written in matrix form:
\begin{equation}\label{EIGEN}
{\cal M}_k{\cal F} ={\cal F}.\end{equation}
This eigenvalue equation  may be easily solved, for instance with
Mathematica. The results are given  in appendix.

Having found the solution under the form of the  vectors ${\cal F}_s$,
with $1 \le s\le \sigma$, with   $\sigma=1/(2J_N+1)$ times  the
degeneracy of the eigenvalue ($\sigma$ is found as a result of the
eigenvalue problem; the value coincide with that given by
\cite{Ikeda}). We have finally a basis $\left( \calY
_{k;M,s}\right)_{M=-k/2..k/2,~ s=1..\sigma}$
for the eigenmodes of $X$ corresponding to the eigenvalue $k$, where
\begin{equation}
\label{FIN}\calY _{k;M,s} \equiv \sum _{\mu=-J_N}^{J_N} ({\cal
F}_s)_\mu~T_{k;M,N~\mu}.
\end{equation}

\section{Application to the spherical spaces}

For each space, the steps are the following:
\begin{itemize}
     \item We write the two generators in the SU(2) $\times$ SU(2)
forms. Since both  are SLARs, this corresponds to two SU(2) matrices
$\gamma _\pm$.  (We thank J. Weeks for providing
generators in suitable form).

     \item We diagonalize the matrix $\gamma _+$.
     \item We deduce the value $J_N$.
     \item We solve the eigenvalue equation (\ref{EIGEN}) in the vector
space of
dimension $2J_N+1$. This gives the vector(s) $\calF$ by their components
     \item Then equ. (\ref{FIN}) gives the eigenmodes of $X$.
     \end{itemize}

\subsection{ Dodecahedral space}

      The two SU(2) generators,    \begin{equation}
    \left(
\begin{array}{c c}
c+iC & \pm i/2 \\
\pm i/2 & c-iC \\
\end{array}
\right);~ c\equiv\cos(\pi/5) ,~C\equiv\cos(2\pi/5),\end{equation}
become,  after diagonalization,
\begin{equation}
\begin{array}{c}
\gamma _{+} =
\left (
\begin{array}{c c}
e^{i\pi/5} & 0 \\
0 & e^{-i\pi/5} \\
\end{array}
\right )
\\
\gamma _{-} =
\left (
\begin{array}{c c}
\cos(\pi/5) + i~\sin(\pi/5)/\sqrt{5} & 2i ~\sin(\pi/5)/\sqrt{5} \\
2i~ \sin(\pi/5)/\sqrt{5} & \cos(\pi/5) - i~\sin(\pi/5)/\sqrt{5} \\
\end{array}
\right )
\\
\end{array}.\end{equation}

We have $N=5$, $J_N= \lfloor k/10\rfloor$.

Equ.(\ref{FIN})  gives the eigenmodes.   We have checked that they
correspond
to those found by \cite{mlrDodec}.

As we mentioned, it may be  more convenient   to express these modes
in the   more widely used basis $\bun$. The
solution of our problem in this basis is, for the dodecahedral space:
\begin{equation}
f_k(x) = \sum_{l,m,m'} i^{m-k-l}(-1)^{m'} f_{k;m'}
(\frac{k}{2}, m'-m; \frac{k}{2}, m' \mid l,m) ~ \mathcal{Y}_{klm},
\end{equation}
where the index $m'$ satisfy the condition (\ref{mcdiag}) for the
dodecahedral space, $l$ vary in $\llc 0,k\lrc$ and $m$ vary in $\llc
-l,l\lrc$.

\subsection{Other multi-connected spherical spaces}

Beside lens and prism spaces,   three other multi-connected
spherical spaces remain. In each one, holonomy groups are generated by
two left-action SO(4) rotations. We  follow the same procedure.

      We give in the table below a list of these spaces with their 2
generators, under their SU(2) form.

\hspace*{-1cm}
\begin{tabular}{l c c}
\\
Tetrahedron &
$\frac{1}{2}\mat{1-i}{-1-i}{1+i}{1+i}$ &
$\frac{1}{2}\mat{1+i}{-1+i}{1+i}{1-i}$ \\
diagonal form & $\mat{e^{i\pi/3}}{0}{0}{e^{-i\pi/3}}$ &
$\frac{1}{\sqrt{3}}\mat{\frac{1}{2}(\sqrt{3}+i)}
{-1-i}{1-i}{\frac{1}{2}(\sqrt{3}-i)}$ \\
\\
\hline
\\
Octahedron & $\mat{e^{-i\pi/2}}{0}{0}{e^{i\pi/2}}$ &
$\frac{1}{\sqrt{2}}\mat{1}{1}{-1}{1}$ \\
\\
\hline
\\
Icosahedron & $\mat{c+i/2}{iC}{iC}{c-i/2}$ &
$\mat{c+iC}{1/2}{-1/2}{c-iC}$ \\
diagonal form & $\mat{e^{i\pi/5}}{0}{0}{e^{-i\pi/5}}$ &
$\mat{c+is}{1/2-ics}{-1/2-ics}{c-is}$ \\
\end{tabular}
with $c:=\cos(\pi/5)$ and $s:=\sin(\pi/5)/\sqrt{5}$.

Computations gives the tables in appendix, where we presented the
first modes for the different spherical spaces. More modes can be
obtain by sending us an email at:
\textrm{cailleri@discovery.saclay.cea.fr}.

Note that the modes for the Icosahedron are the same than those of the Dodecahedron, as expected; they are however given here in a different frame.

%%%%%%%%%%%%%%%%%%%%%%%%%%%%%%%%%%%
%
\section{Conclusion}
%        ----------
%%%%%%%%%%%%%%%%%%%%%%%%%%%%%%%%%%%

The introduction of the  basis $\bdeux$ of Wigner functions provided
the explicit
behavior  of the modes of $\st$,   under the  rotations of
SO(4). This allowed to write an invariance condition of these modes
under any rotation, and under the holonomy group $\Gamma$ of a
spherical space
$\st/\Gamma$.
The solutions of the corresponding equations give the vector spaces of
the eigenmodes of  $\st/\Gamma$. These modes are calculated
explicitly, for the first values of $k$, for the spherical spaces, in
the basis
$\bdeux$. We give the transformation relations to express them in the
more usual basis $\bun$ of hyperspherical harmonics.

This allows to calculate the  power angular  spectrum of the CMB
fluctuations,
as predicted by the cosmological model with Poincar\'e dodecahedral
space, up to arbitrary multipoles $k$. This work is in progress. It
will allow to check if  the predictions of \cite{Luminet}~hold up to
higher orders. In any case, the confrontation of the predicted power
spectrum with observations may not offer enough discriminating
power. Thus, we intend to calculate the cross correlations predicted
from this model. They are zero for Gaussian isotropic fluctuations,
and thus allow more specific tests of cosmic topology. This work is
also in progress.

The present results   allow to   extend these calculations  to the
remaining spherical spaces, a work also in progress. Hopefully, this
will provide a definitive
test  of the case of   a multi-connected space with positive
curvature, at least for a range of scales comparable with that of the
last scattering surface.

%%%%%%%%%%%%%%%%%%%%%%%%%%%%%%%%
%

%%%%%%%%%%%%%%%%%%%%%%
%                    %
%      Appendix      %
%                    %
%%%%%%%%%%%%%%%%%%%%%%
\newpage
\appendix

\begin{center}
\section*{Table of eigenfunctions for the dodecahedral space}
\end{center}

Eigenmodes of dodecahedral space corresponds to the basis of functions
$(f_{k}(x))_{m\in\llc -k/2;k/2\lrc}$, where
$f_k=\sum_{m,m'}f_{k;m'}T_{k;m,m'}$. In this table is indicated the
allowed $k$ (those where $D_k$ admits $1$ as an eigenvalue) and the
corresponding $f_{k;m'}$, for $m'=0\pmod{5}$ (for others values of
$m'$, $f_{k;m'}=0$).

\vspace*{0.7cm}

\begin{tabular}{c c c c}
\textbf{k=12} & & & \\
0.529150262\,i & 0.663324958 &
0.529150262\,i & \\
\hline \\
\textbf{k=20} & & & \\
-0.3158058475\,i & 0.578273291 &
0.362950869\,i & 0.578273291 \\
-0.3158058475\,i & & & \\
\hline \\
\textbf{k=24} & & & \\
-0.486949689 & 0.3025227264\,i &
0.585422924 & 0.3025227264\,i \\
  -0.486949689 & & & \\
\hline \\
\textbf{k=30} & & & \\
-0.2829840984& 0.391305507\,i & -0.516526862 & 0 \\
0.516526862 & -0.391305507\,i & 0.2829840984 & \\
\hline \\
\textbf{k=32} & & & \\
-0.329930901\,i & 0.425449096 & -0.331083071\,i & 0.448380790 \\
-0.331083071\,i & 0.425449096 & -0.329930901\,i & \\
\\
\end{tabular}

   For this space and others, we have calculated more modes, and with an
higher precision. As indicated in the text, these modes could be asked
at: \textrm{cailleri@discovery.saclay.cea.fr}.

%%%%%%%%%%%%%%%%%%%%%%%%%%%%%%%%%%%%%%%%%%%%%%%%%%
\newpage

\begin{center}
\section*{Table of eigenfunctions for the tetrahedral space}
\end{center}

   For this space, we present modes under the form: $f_{k;m}$, with the
notation used above, where $m=0\pmod{3}$. For $k=12$, we  can observe
that modes are degenerated and we thus have 2 series of $f_{k;m}$.
\vspace*{0.7cm}

\begin{tabular}{c c}
\textbf{k=6} & \\
0.333333333 - 0.333333333\,i & 0.745355992 \\
-0.333333333 - 0.333333333\,i \\
\hline \\
\textbf{k=8} & \\
0.608580619 & -0.360041149 - 0.360041149\,i \\
-0.608580619\,i \\
\hline \\
\textbf{k=12} & \\
-0.00323332162 + 0.00323332162 \,i & 0.738332205 \\
0.2666917602 + 0.2666917602 \,i & 0.2828693156\,i \\
  -0.341013640 + 0.341013640 \,i \\
\hline \\
0.354280119 + 0.354280119 \,i & -0.0861446455 \,i \\
-0.353087976 + 0.353087976 \,i & 0.689157164 \\
-0.0929259330 - 0.0929259330\,i \\
\hline \\
\textbf{k=14} & \\
0.491351820\,i & -0.1888525745 + 0.1888525745\,i \\
0.611952283 & 0.1888525745 + 0.1888525745\,i \\
-0.491351820\,i \\
\\
\end{tabular}

%%%%%%%%%%%%%%%%%%%%%%%%%%%%%%%%%%%%%%%%%%%%%%%%%%%%%%

\newpage
\begin{center}
\section*{Table of eigenfunctions for the octahedral space}
\end{center}

   For this space, we present modes under the form: $f_{k;m}$, with the
notation used above, where $m=0\pmod{2}$.
\vspace*{0.7cm}

\begin{tabular}{c c c c}
\textbf{k=4} & & & \\
-0.612372435 & 0.500000000 & -0.612372435 & \\
\hline \\
\textbf{k=8} & & & \\
-0.379649579 & -0.1447862189 & -0.818416944 & -0.1447862189 \\
-0.379649579 & & & \\
\hline \\
-0.530330085 & 0.3118047822 & -0.493006648 & 0.3118047822 \\
-0.530330085 & \\
\hline \\
\textbf{k=10} & & & \\
-0.353553390 & -0.612372435 & 0 & 0.612372435 \\
  0.353553390 & \\
\hline \\
\textbf{k=12} & & & \\
0.507752400 & -0.1250000000 & 0.342326598 & -0.467707173 \\
0.342326598 & -0.1250000000 & 0.507752400 & \\
\\
0.1230304012 & -0.672015232 & 0.0829470795 & 0.2296903537 \\
  0.0829470795 & -0.672015232 & 0.1230304012 & \\
\\
\end{tabular}

%%%%%%%%%%%%%%%%%%%%%%%%%%%%%%%%%%%%%%%%%%%%%%%%%%%%%%

\newpage
\begin{center}
\section*{Table of eigenfunctions for the icosahedral space}
\end{center}

   For this space, we present modes under the form: $f_{k;m}$, with the
notation used above, where $m=0\pmod{5}$.
\vspace*{0.7cm}

\begin{tabular}{c c}
\textbf{k=12} & \\
0.1882530268 + 0.3999418963\,i & 0.6533724434 \\
  -0.2617387416 + 0.5560616500\,i \\
\hline \\
\textbf{k=20} & \\
-0.2115446757 - 0.0542397647\,i & 0.3004753102 - 0.3633777752\,i \\
  0.1486544684 + 0.3158151081\,i & 0.655649414\\
  0.1798126817 - 0.3820104577\,i \\
\hline \\
\textbf{k=24} & \\
-0.0622550053 - 0.3255605557\,i & -0.2352778614 - 0.0603249209\,i \\
-0.3529784282 + 0.4268720642\,i & -0.1437372075 - 0.3053684305\,i \\
0.6406372434 & \\
\hline \\
\textbf{k=30} & \\
0.02978235622 + 0.1557458776\,i & 0.2506842548 + 0.0642750991\,i  \\
-0.2564612869 + 0.3101497151\,i & 0.0002877519 + 0.0006113259\,i \\
0.560360227 & 0.2128816958 - 0.4522652865\,i \\
0.2718985718 + 0.3288186907\,i & \\
\\
\end{tabular}

\end{document}